# AEcroscopyWave: Towards Self-Driving Characterization Platforms for Agentic AI

Yongtao Liu,[1,*] Jawad Chowdhury,[1] Ganesh Narasimha,[1] Ralph Bulanadi,[1] Liam Collins,[1] Ruben Millan Solsona,[1] Marti Checa,[1] Asraful Haque,[1] Sumner B. Harris,[1] Stephen Jesse,[1] Rama Vasudevan[1,**]

[1] Center for Nanophase Materials Sciences, Oak Ridge National Laboratory, Oak Ridge, TN 37830, USA

*liuy3@ornl.gov
**vasudevanrk@ornl.gov

**Abstract**
The characterization of electronic materials has traditionally been stratified into two distinct regimens: industry-scale automated systems to inspect materials for defects and ensure quality (such as in the semiconductor industry), and highly customized, operator-driven systems requiring human experts. The former offers high throughput but limited flexibility, whereas the latter is heavily bandwidth-limited but provides research-grade discovery capabilities. Recent advances in "self-driving" characterization tools offer the potential to bridge the two stratified regimes, by the creation of application program interfaces (APIs) that can control hardware, and the integration of AI methods to incorporate autonomy into the process. Here, we discuss our latest developments in AEcroscopyWave, a custom-built characterization platform for the agentic-AI era, that provides unified control of scanning probe microscopes with programmable peripheral instrumentation, highlighting the design choices that are necessary for maximizing the capability of the system and the ease of use for both human and AI agents. The benefits of making heterogeneous scientific instruments accessible, composable and usable by agents is demonstrated by test cases.

## Introduction

Artificial intelligence (AI)/ machine learning (ML) is playing an increasingly important role in scientific research, spanning from data analysis to autonomous decision-making and active control of experiments.[1-5] More recently, the role of AI/ML has expanded beyond passive data analysis toward active participation in the experimental process itself. This shift has led to growing interest in autonomous experimentation (AE) and/or self-driving laboratories (SDLs),[6-10] where AI/ML iteratively analyzes experimental results, determines subsequent experimental conditions, and controls instrumentation in real time with minimal human intervention. Such approaches have demonstrated substantial impact in areas including battery materials,[11, 12] photovoltaic materials,[13-15] thin-film synthesis,[16-22] materials characterization,[23-29] and polymers.[30-32] By combining automated instrumentation and AI/ML algorithms within expert-designed workflows, autonomous platforms can explore complex experimental spaces far more efficiently than traditional manual, trial-and-error approaches.

Instrument automation is a prerequisite for autonomous experimentation. While scripting and workflow-control environments have long enabled predefined measurement routines, control of characterization instruments is increasingly being exposed through machine-accessible interfaces that allow their operation to be coordinated from high-level programming environments.[19, 33-35] These interfaces may take the form of vendor application programming interfaces (APIs), open-source control packages,[36-39] or custom software layers built around lower-level instrument commands.[33, 35] By connecting instrument control with on-the-fly analysis and decision-making, such software architectures enable active-learning workflows in which methods such as Bayesian optimization can guide experiments through carefully defined parameter spaces with reduced human intervention in both synthesis and characterization[6, 8].

Despite the gains, this requirement on only navigating predefined parameter spaces limits our overall capacity to creatively adapt and modify experimental workflows *ad-hoc* with changing experimental objectives in real-time. Moreover, it comes at a time when progress from agentic AI systems is accelerating, with systems now capable of some forms of reasoning,[40, 41] as well as autonomous hypothesis generation.[42, 43] Agentic AI systems are those which we define as AI systems which can plan, iterate, create, and execute tasks until some objective is completed.[44-46] However, current autonomous lab APIs typically do not offer services that are powerful for both humans and AI agents, given their requirements and capabilities are quite different. That is, traditional software favored ease of use over customizability and flexibility, to enable users to more easily e.g., select from lists of allowed operations rather than enable new modalities to be programmed even if the hardware could implement it. However, fully realizing the benefits of agentic AI systems means that future experimental platforms must support far richer and less constrained interactions between AI and instrumentation.[47-50] In these cases, the role of AI is no longer limited to sampling within predefined parameter spaces (e.g., BO based workflows). Instead, AI systems are expected to generate entirely new experimental conditions and/or strategies that may differ substantially from prior human-defined conditions. For example, microscopy and spectroscopy often rely on specially designed excitation signals to probe time-dependent and/or dynamical material responses; thus, different excitation waveforms, or sequences of waveforms could be used to selectively access distinct physical mechanisms within materials.

This emerging need motivates the development of experimental platforms that allow AI systems to directly program and modify experimental protocols,[47, 51] and builds on the concept of 'autoresearch' recently espoused by Karpathy[52] and implemented in the case of a chemistry self-driving labs (SDL) by Aspuru-Guzik.[53]. Here, we discuss the development of the automated microscopy platform AEcroscopy (short for automated experiments for microscopy) to illustrate our perspective on instrument automation in the era of agentic AI.[33] Specifically, we discuss the evolution of the AEcroscopy system from version 1 to version 2 (*i.e.*, AEcroscopyWave), the version 2 includes a specific distributed agent model that separates planning from execution, a model context protocol (MCP) server for agent-based operations, human approval gates, and experimental and waveform planning, as shown in Figure 1. Hereafter, AEcroscopy v1 will be referred to as AEcroscopy, and AEcroscopy v2 as AEcroscopyWave. While AEcroscopy primarily focused on automating experimental execution within predefined workflows, AEcroscopyWave introduces substantially greater flexibility through programmable and software-defined instrument control. In particular, the ability to dynamically generate and upload arbitrary excitation waveforms enables AI systems to move beyond conventional parameter optimization toward creative experimental design. We envision such capabilities as essential building blocks for next-generation autonomous microscopy platforms, where AI agents are capable not only of executing experiments autonomously, but also of inventing, refining, and adapting experimental strategies in response to evolving scientific observations, whilst still maintaining human oversight.

**AEcroscopy**

The key advantage of modern-day microscopy for nanoscale characterization is that it enables correlation of microstructural features of samples with functional properties, by mapping structure from imaging studies to spectral features (from spectroscopic measurements conducted at the same location), which can be correlated with material properties, thereby enabling the deduction of structure-function relationships. These relationships are often hidden in ensemble-averaged measurements at the bulk level. Further, by integrating imaging with time-resolved spectroscopy, one can determine how these local structural variations influence dynamic processes and energy dissipation pathways. This combined approach, simultaneously accesses structural and dynamic information, enables deeper insight into the coupling between morphology and functionality, provides a more complete picture of material functionality, and can guide the design of materials with optimized performance.

AEcroscopy is a combined hardware-software platform to control a variety of microscopy tools using python, including atomic force microscopes (AFM), scanning transmission electron microscopes (STEM), and scanning tunneling microscopes (STM).[33] A central contribution of AEcroscopy was the ability to automatically control probe motion and perform spectroscopy measurements without continuous human intervention. AEcroscopy provides direct control over probe position, automating this workflow by providing programmatic control of the microscope, which can acquire images, analyze them, select locations of interest, move the tip automatically, and trigger spectroscopy measurements.

AEcroscopy also enabled automated high-throughput spectroscopy measurements. Instead of manually measuring only a few selected points, the platform can automatically acquire spectroscopy data across a list of predefined spatial locations. By directly integrating image analysis algorithms with microscope control, AEcroscopy creates a closed-loop experimental workflow targeting spectroscopy measurements. For example, ML models can identify features of interest from microscopy images, after which AEcroscopy automatically positions the probe at the selected locations and performs spectroscopy measurements.[54-57] This integration of ML into the workflow enables more advanced and intelligent characterization strategies, e.g., an AFM image can first be acquired to identify regions or features of interest, an ML model is then applied to determine optimal locations for spectroscopy measurements, AEcroscopy subsequently drives the tip to each selected coordinate and performs local spectroscopy automatically. Importantly, the entire process, including image acquisition, feature identification, point selection, tip positioning, and spectroscopy measurement, is executed autonomously without manual intervention.

AEcroscopy also supports adaptive and autonomous experimentation, where the outcome of previous spectroscopy measurements guides the selection of future measurement locations through BO or related active-learning algorithms. For example, active learning approaches have been employed to guide spectroscopy sampling to accelerate structure–property analysis;[58-60] novelty-driven[61-64] and curiosity-driven learning strategies[65, 66] have been developed to autonomously explore spectroscopic responses containing previously unobserved or information-rich behaviors.

In addition to spectroscopy measurements, AEcroscopy enables automated nanoscale material manipulation. For example, voltage pulses can be applied to locally switch ferroelectric polarization states or move domain walls. Researchers can systematically investigate how pulse parameters influence polarization switching mechanisms,[67, 68] enabling deeper understanding of material behavior. ML approaches such as reinforcement learning (RL) can be trained to autonomously create user-defined nanoscale structures through adaptive pulse control.[69, 70]

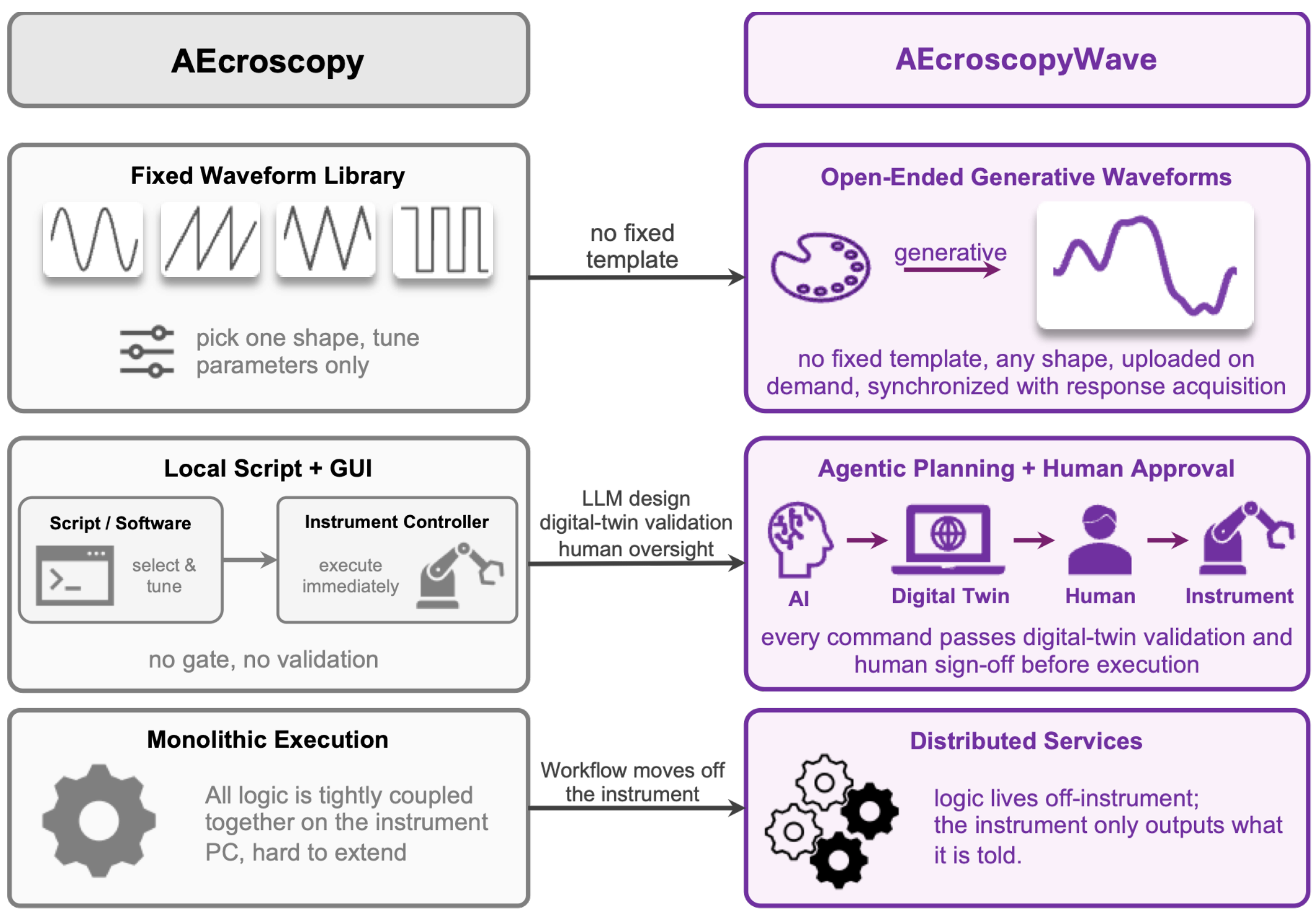

**Figure 1.** Illustration of the transition from AEcroscopy to AEcroscopyWave. Waveform generation shifts from a fixed library of predefined shapes (with only parameter tuning) to open-ended generative waveforms created on demand and synchronized with response acquisition; execution shifts from immediate local script/GUI commands with no validation to an agentic workflow where every command passes AI planning, digital-twin validation, and human approval before execution; and the architecture shifts from monolithic logic on the instrument PC to distributed off-instrument services, with the instrument only executing what it is told.

## AEcroscopyWave

In the above mentioned instances describing both spectroscopy and manipulation experiments, AEcroscopy does not support the creation of entirely new probing stimulus, e.g., excitation waveforms. Instead, waveforms are typically selected from a predefined set, with researchers only able to adjust parameters such as amplitude, frequency, or duration. This paradigm aligns well with traditional BO or RL, where ML models optimize within a fixed action or parameter space. As AI evolves toward generative capabilities, where AI can autonomously design and create novel waveforms rather than simply choose from existing options, automated experimental platforms must also evolve to support this new level of flexibility. Upgrading automation systems to enable on-the-fly waveform generation and deployment (as demonstrated in Figure 2) will allow generative AI and autonomous experimentation platforms to work synergistically, further accelerating scientific discovery and materials research.

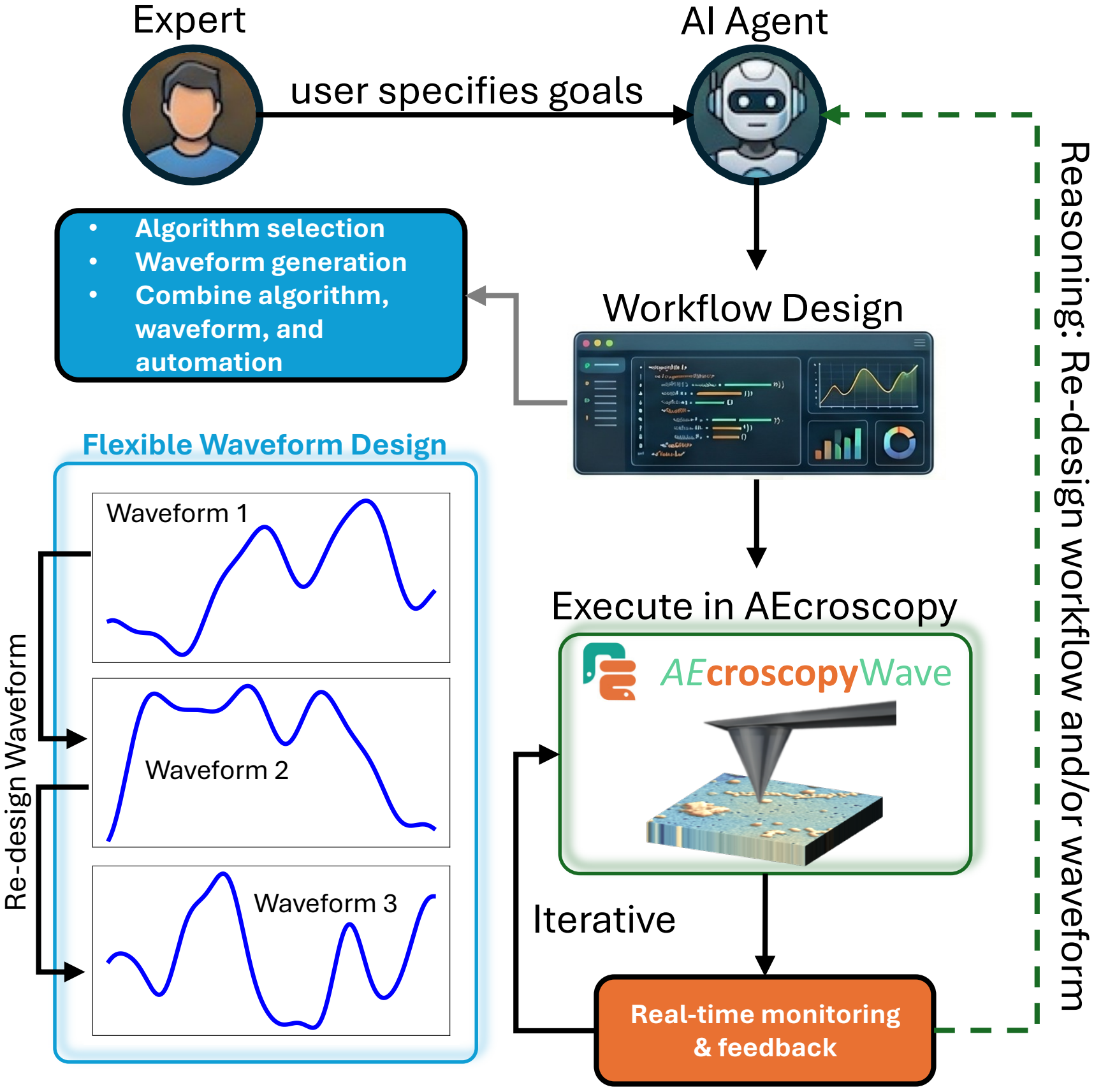

**Figure 2.** Illustration of AEcroscopyWave for agentic AI-driven autonomous experimentation, where human experts define scientific objectives while AI agents dynamically design new experiments and/or arbitrary excitation waveforms based on real-time experimental observations and feedback.

This emerging need motivated the upgrade of the AEcroscopy to AEcroscopyWave, where we introduced the capability to dynamically upload and output entirely user-defined excitation waveforms. To facilitate the transition to the agentic AI, we made several key changes to the architecture of AEcroscopy. First, the original AEcroscopy utilized a purely local execution on the instrument, with scripts controlling instrument functions, and more complex Labview-based GUIs that could be used to select from predefined waveforms to send to the controller to accomplish spectroscopy and imaging workflows. These choices were made specifically for reducing the complexity of programming the most common types of workflows encountered. In AEcroscopyWave, we instead use a distributed agent-based architecture with planning and approval gates, and a much simpler LabVIEW-based instrument control interface. Together, these capabilities establish AEcroscopyWave as a software-defined experimental platform that enables AI systems to dynamically design, execute, and adapt microscopy experiments in real time. Below, we will discuss the architectural pieces of AEcroscopyWave.

**Distributed Agent-based architecture**

AEcroscopyWave replaces the original AEcroscopy's monolithic design with a distributed server-client model, that separates code planning, human oversight and approval, and execution, onto different computers connected over a network, as shown in Figure 3. Instruments are connected to a cloud server, where a persistent afm-server is running using FastAPI, and serves to coordinate jobs. This server maintains an SQLite database for the available instruments and a job queue, accepts job submissions from users, and then dispatches them to instrument agents (clients). Meanwhile, the clients (afm-agent) are run on each instrument, and register with the server the instrument details, the types of jobs they can accept, and poll the server at a set interval for any tasks needed to be executed. When there is a queued job that needs to be run on that instrument, the agent takes the job, executes it, then posts the result back to the server using the REST API. We also provide a streamlit dashboard, which includes an large language model (LLM)-assisted experimental planner, an approval queue, a results viewer, and a named workflow library, among other features.

One important distinction made here is that the agent cannot simply run any job directly on the instrument: when it receives a job from the server, it cross-checks that the job matches one of the registered hardware handlers (such as 'set_setpoint', 'move_tip', etc.), or whether it matches a multi-step workflow in the mode context protocol (MCP) workflow registry, or, in the case of custom scripts, whether the custom script has been human-approved. This ensures that every code has been checked prior to execution.

Furthermore, we provide a 'digital twin' interface for some available backends (such as the DriveAFM and Cypher tools), to validate custom scripts before running on the instrument. We note that 'digital twin' in this instance refers to simply a virtual microscope that simulates the control execution aspects rather than the strict National Academy definition; however, efforts are ongoing towards realizing true digital twins. All customized scripts must first pass through the digital twin without errors before they can be approved to pass through to the job queue, for syntactic verification.

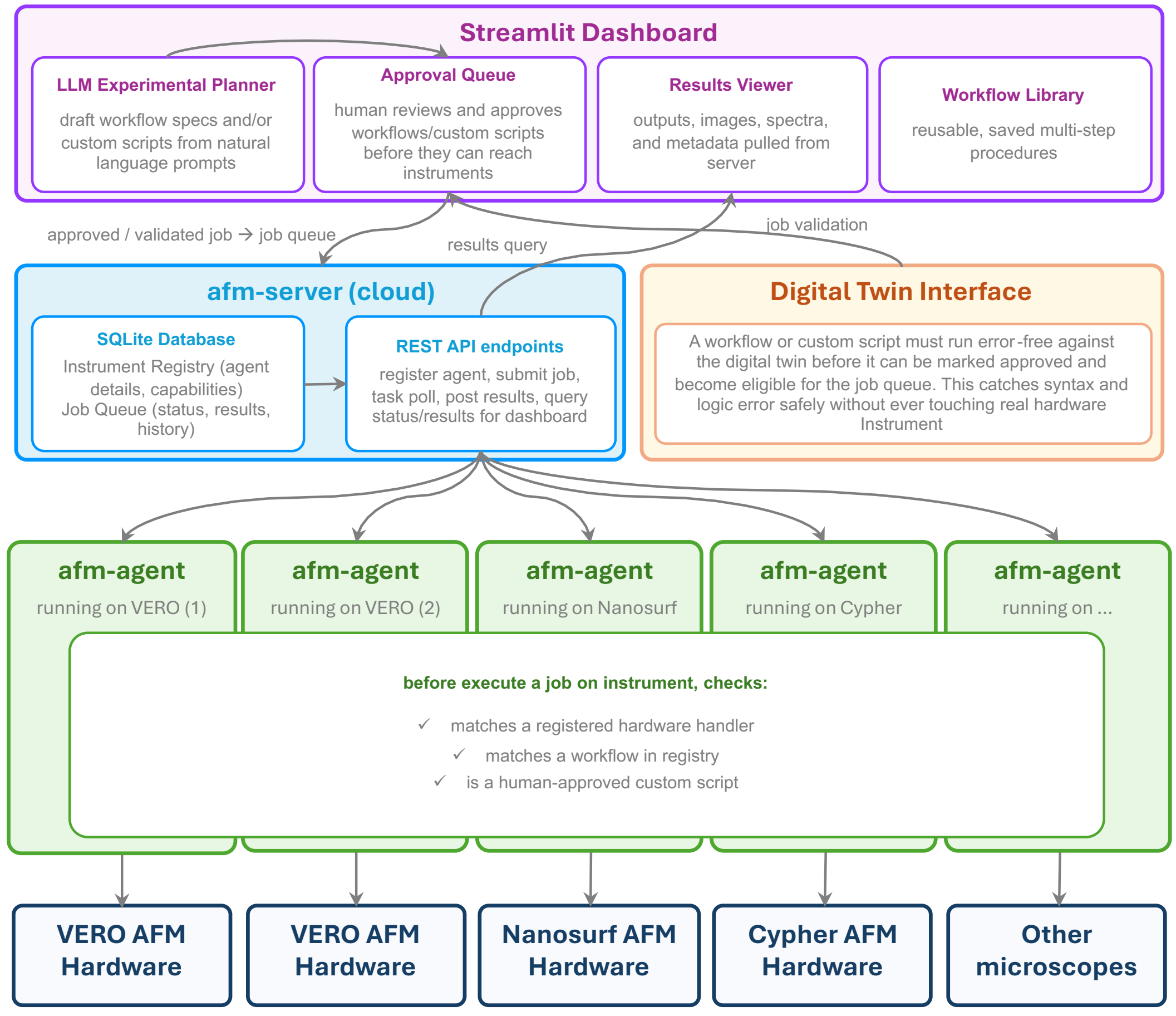

Figure 3. Distributed agent-based architecture.

## MCP Tool Server

To make AEcroscopyWave “agent-native”, a key design was the creation of an MCP tool server to expose the microscope functions to the agent in an easy-to-understand manner. In AEcroscopy, writing scripts required human experts to read through modular code blocks of abstract classes, find the right functions, and use the right argument types for specific backends, and chain them together to form experimental workflows. However, agents benefit from a flat, discoverable tool layer with consistent calling conventions. For this, AEcroscopyWave utilizes the MCP via the FastMCP framework, thereby making any LLM that can utilize MCP tools capable of calling AEcroscopyWave functions. Tools are organized into groups, including a special group called ‘workflows’, which is used to list, inspect and execute available ‘workflows’, i.e., predefined experimental routines which users often run. Notably, previously validated workflows or commonly used experimental procedures can be added to the MCP workflow registry. Users can then invoke these workflows directly by name when needed, allowing the agent to execute pre-approved, standardized sequences of actions without requiring additional code generation or review, thereby improving both efficiency and operational safety.

**Experiment Planning and Waveform Generation**

Two key agentic features in AEcroscopyWave are experiment planning with LLM assistance, as well as customized waveform generation. The LLM Planner uses the MCP tool catalog to generate scripts from natural language descriptions input by user. The user utilizes the streamlit interface to enter their description of the experiment into the prompt window, and we utilized an Ollama server and gpt-oss120b model along with a system prompt containing the tool catalog (names, descriptions, parameter signatures), workflow examples, and a response-format contract that constrains the LLM to produce a python run() function, using either registered tools only, or if not possible, with newly generated code that must be human-approved. The modular design of AEcroscopyWave also enables integration with a variety of commercial LLM platforms.

In AEcroscopyWave, autonomous experimentation is no longer constrained by predefined excitation waveforms or fixed parameter spaces. Instead, once a user specifies a scientific objective or research goal, the AI agent (with instrument capabilities exposed through MCP-compatible tools) can autonomously design both the experimental workflow and the excitation strategies required to achieve that objective. For this purpose, AEcroscopyWave supports fully programmable and arbitrary waveform generation. Rather than selecting from a predefined library of excitation functions, AI agents can dynamically generate, modify, and refine excitation waveforms during experiments, including waveform structures that may not correspond to previously known or human-designed forms.

This is achieved by a three-layer waveform architecture. The WaveGenerator module provides a library of waveform primitives, such as square waves, triangular waves, sine waves, chirps, etc., as well as more advanced spectroscopic waveforms. Users and AI agents can directly use these functions to generate a tremendous assortment of custom waveforms, and a named waveform registry provides discoverability. New waveforms can also be easily created using these primitives, and registered for re-use within experimental campaigns. At the third level, there exist MCP tools to expose waveform generation as agent-callable operations, to enable the generation, upload and execution of the generated waveforms on the microscope instrument.

**Hardware-Software Interface**

To interface with experimental hardware, AEcroscopyWave includes control modules for both data acquisition systems and scientific instruments. The WaveVI module provides an interface to a LabVIEW-controlled National Instruments DAQ system, enabling configuration of acquisition parameters, waveform upload, hardware-triggered signal output, and analog input data collection. As a complement to WaveVI, AEcroscopyWave also includes alternative interfaces to waveform generation hardware including Liquid Instruments and Zurich Instruments for users who prefer commercial waveform generation and execution hardware, providing an alternative platform for waveform deployment and synchronized data acquisition. For microscope control, AEcroscopyWave provides unified interfaces for several common microscopy platforms, including Asylum Research, Nanonis, and Nanosurf systems. These interfaces support a broad range of instrument operations, including scan parameter configuration, imaging control, tip engagement and motion, crosspoint routing, environmental monitoring, and execution of low-level instrument commands. Additional waveform execution backends and microscope control interfaces will be continuously incorporated into future releases, further expanding hardware compatibility and supporting a broader range of experimental platforms.

## Case Study

As an example of the use of AEcroscopyWave for flexible waveform generation and execution, we show the generation of novel waveforms for switching in ~200nm-thick $BaTiO_3$ (BTO) thin film deposited on (001) $SrTiO_3$ with a thin $La0_{.67}Sr_{0.33}MnO_3$ bottom electrode, grown by pulsed laser deposition (PLD). We designed a pulse-train piezoresponse force microscopy (PFM) experiments to test how prior electrical conditioning and post-conditioning delay can influence ferroelectric switching. The general idea of the experiment is to understand the degree to which surface injected charges are influencing the polarization switching process (Fig. 4[a]). In an ideal fully screened case, there should be no dependence on conditioning pulses and delay times; however, BTO thin films are well known for surface ionics, and a variety of surface electrochemical and surface-bulk coupled processes may affect ferroelectric switching dynamics.

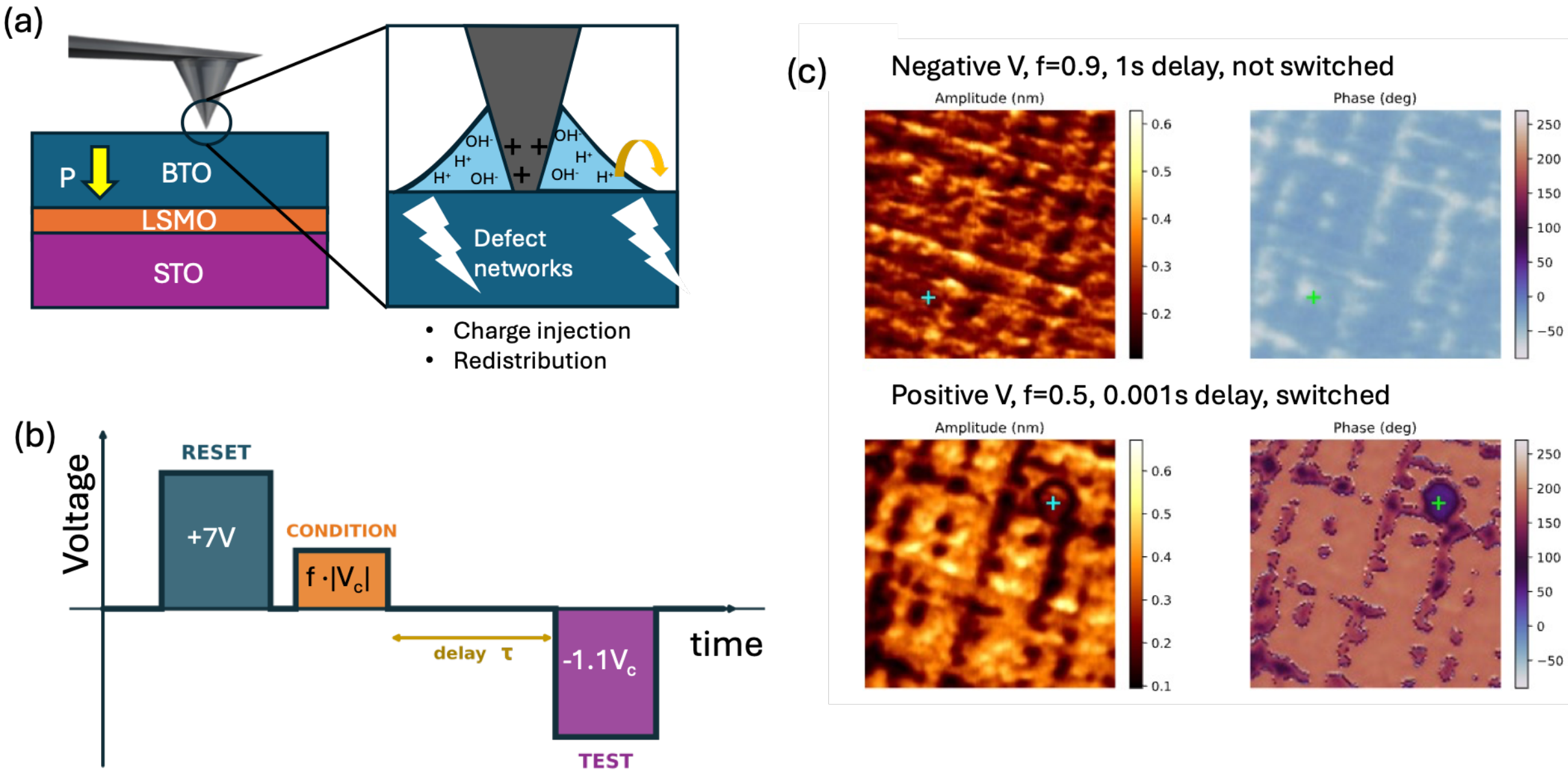

**Figure 4:** Pulse train experiment. (a) A schematic of our setup, where a conductive probe is in contact with the BTO surface. Inset, a meniscus layer around the tip, and expected charge redistribution caused by bias application to the tip. Defect networks within the sample can couple with surface ionics leading to different relaxation times and influence ferroelectric screening and switching processes. (b) Voltage waveform setup, with a reset pulse, conditioning pulse, delay time and a test pulse. (c) Data from the experiment, showing one instance where the sample was unswitched (upper), and once instance where the sample was switched (lower). Images are PFM amplitude and phase, 1 μm × 1 μm size.

The design was a randomized full factorial screen across three conditioning amplitudes and six delays. Each condition used a strong reset pulse, a subthreshold same-polarity conditioning pulse, a controlled delay, and a near-threshold opposite-polarity test pulse [Fig. 4(b)], followed by a full PFM frame. The main idea is to determine if there is a dependence on the resultant domain size on the delay time and the amplitude of the conditioning pulse. A key strength of this setup is the flexibility of the input excitation waveform: the experiment can independently tune pulse polarity, conditioning fraction, timing delay, reset/test amplitudes, and waveform upload structure, making it adaptable for probing many switching, relaxation, and memory effects without redesigning the whole protocol. Note that the number of conditioning amplitude and delay in this experiment can be customized by users, the platform supports any amounts of conditions.

The results showed switching in 42 of 72 conditions, for an overall switching probability of 0.58. An example of a switched case and an unswitched case is shown in Fig. 4(c). We analyzed the data by computing the difference maps between the previous and next PFM images, as shown in Fig. 5(a), to extract the written domain radius as a function of the waveform parameters and plot them for each individual pulse location in Fig. 5(b). The experiment was performed at two different sites per image, for different images for a total of four distinct points. The strongest qualitative trend was a non-monotonic conditioning effect: a conditioning amplitude fraction f=0.5 produced the highest switching probability, 0.75, compared with 0.54 for f=0 and 0.46 for f=0.9, and also had the largest mean domain area among all conditions. However, these trends were marginal rather than statistically significant, with conditioning fraction giving p=0.117 for switching probability and p=0.078 for area across all conditions.

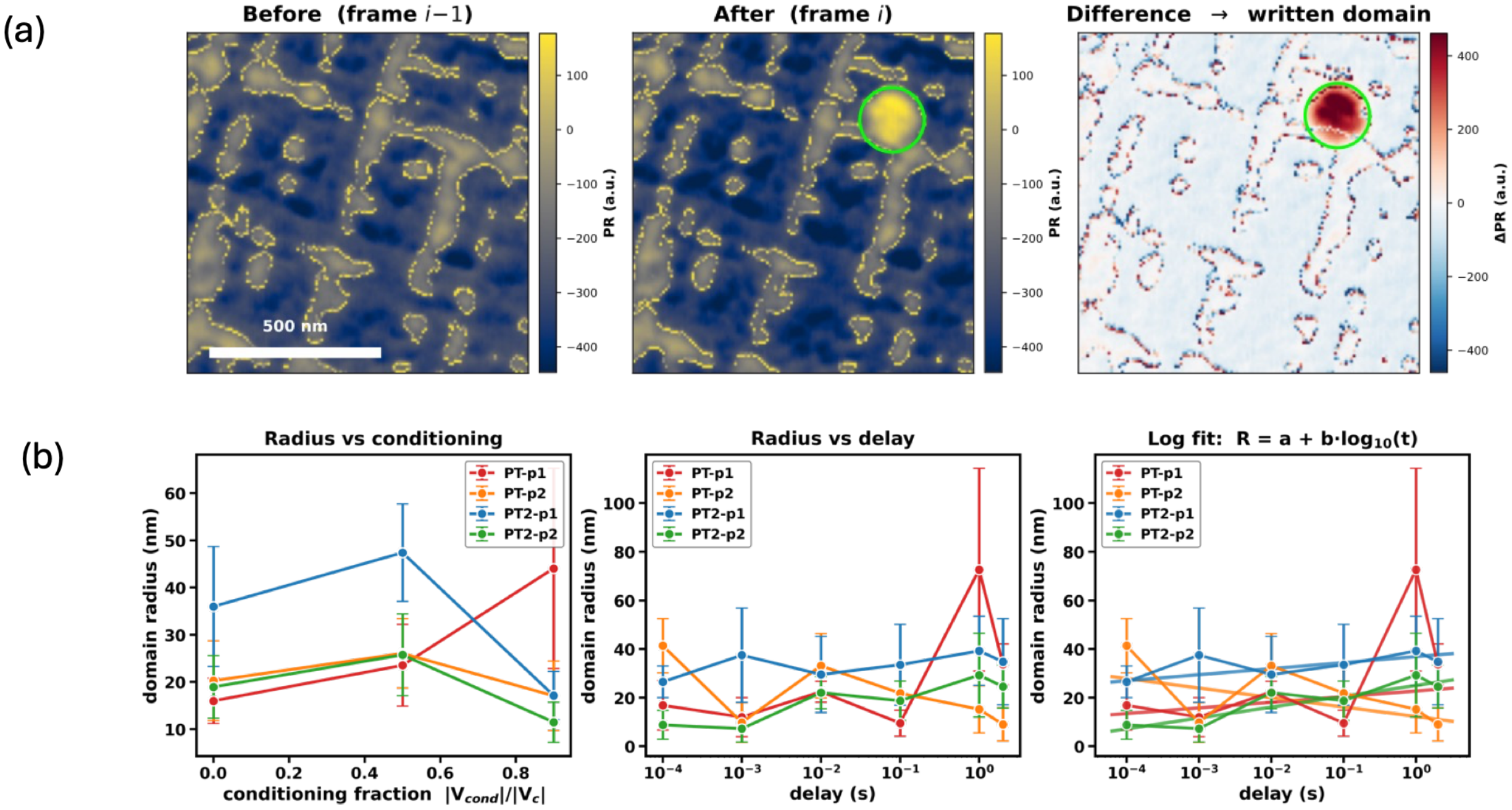

**Figure 5:** Processing and analysis of spectroscopic imaging dataset on BTO. (a) PFM images before and after the pulses are used to calculate the difference image, and the switched domain radius area in the location around the site where the pulse was applied is calculated. This strategy was used to calculate the radius as a function of the pulse conditioning fraction and delay times for the experiments. Results for all variables, for the four different positions (PT1 is one image location, PT2 is a second location on the same sample, and pt1-2 refer to different points on the same image) are shown in (b).

The delay did not affect switching probability, but among switched domains there was a weak positive trend between delay and domain size, suggesting possible relaxation- or retention-related growth after conditioning, though this also remained below significance. Overall, the results provide suggestive evidence that intermediate subthreshold conditioning may enhance switching more than either no conditioning or stronger subthreshold conditioning. Additionally, there was a strong point-to-point variability in the results, which is directly evident from the PFM images that may point towards some form of extended defect network present within the film. Fitting the time-dependence reveals logarithmic decay and is consistent with extended defect (dislocation) networks coupled to mobile surface ionics (protons) (Fig 5c). This study highlights how customized waveforms in AEcroscopyWave can provide unique insights into electronic and semiconductor materials that are otherwise difficult to capture without specialized setups.

**Summary**

As characterization instruments continue to become more automated, the need to design systems that can be amenable to both human and AI-agent operations is critical if we are to maximize their utility for autonomous laboratories. Whereas past software control paradigms focused on ease of use, new agentic platforms can focus on flexibility, hardware utilization, safety, and human planning and oversight. AEcroscopyWave describes an advancement of microscopy control in an agentic AI era, showcasing benefits of making heterogeneous scientific instruments accessible, composable and usable by agents. By enabling customized experimental planning, waveform generation, code validation and execution via MCP, our autonomous microscopy framework offers a vision for future distributed coordination of microscopy experiments that is composable and scalable for human-AI collaborations in an agentic era.

**Acknowledgement**

The agentic planning, distributed architecture and digital twin integrations was demonstrated and validated with support by the Center for 3D Ferroelectric Microelectronics Manufacturing (3DFeM2), an Energy Frontier Research Center funded by the US Department of Energy, Office of Science, Office of Basic Energy Sciences Energy Frontier Research Centers program under award DE-SC0021118. The MCP agent work and experimental work was supported by the Center for Nanophase Materials Sciences (CNMS), which is a US Department of Energy, Office of Science User Facility at Oak Ridge National Laboratory.

**Data and Code Availability**

The user documentation for AEcroscopyWave is publicly accessible at https://yongtaoliu.github.io/aecroscopy.pyae/welcome_intro.html .

**Conflicts of Interest**

The authors declare no conflicts of interest.

**Authors Contribution**

Y.L. and R.V. conceived the idea. Y.L. developed the AEcroscopyWave Python library to enable communication with the hardware and constructed the MCP server. R.V. developed the experiment planner, digital twin, scheduler, and agent. L.C. and R.M.S integrated the instances for Liquid Instruments and Nanosurf DriveAFM. G.N. developed the instance for Nanonis STM. A.H. and S.B.H. synthesized BTO thin film. S.J. built the LabVIEW interface. All authors contributed to the manuscript.